\documentclass[preprint2,twoside,dvips]{aastex}
\usepackage{graphicx}
\newcommand{\gae}{\mathrel{>\kern-1.0em\lower0.9ex         
\hbox{$\sim$}}}

\begin{document}        
\title{THERMAL AND NONTHERMAL X-RAY EMISSION IN SNR RCW 86}
\author{Kazimierz J.~Borkowski}
\affil{Department of Physics, North Carolina State         
University, Raleigh, NC 27695}
\email{kborkow@unity.ncsu.edu}        
\author{Jeonghee Rho}
\affil{IPAC/California Institute of Technology, Pasadena, CA 91125}
\email{rho@ipac.caltech.edu}
\author{Stephen P.~Reynolds, and Kristy K.~Dyer}
\affil{Department of Physics, North Carolina State         
University, Raleigh, NC 27695}
\email{Stephen\_Reynolds@ncsu.edu, Kristy\_Dyer@ncsu.edu}

\begin{abstract}
Supernova remnants may exhibit both thermal and nonthermal X-ray emission.
Such remnants can be distinguished by the weakness of their X-ray lines,
because of the presence of a strong nonthermal X-ray continuum. RCW 86 is
a remnant with weak lines, resulting in low and peculiar abundances when
thermal models alone are used to interpret its X-ray spectrum. This 
indicates the presence of a strong nonthermal synchrotron continuum.

We analyze ASCA X-ray spectra of RCW 86 with the help of both
nonequilibrium ionization thermal models and nonthermal synchrotron
models.  A two-temperature thermal model and a simple
nonthermal model with an exponential cutoff (plus interstellar
absorption) give reasonable results. We obtain blast wave velocity of
800 km s$^{-1}$, the shock ionization age of $1-3
\times 10^{11}$ cm$^{-3}$ s, and the break in nonthermal spectra at
$2-4 \times 10^{16}$ Hz. The
strength of nonthermal continuum correlates well with the radio
brightness in the bright SW section of the remnant. This is convincing
evidence for X-ray synchrotron emission in RCW 86.
\end{abstract}        
        
\section{RCW 86 AND ITS X-RAY SPECTRUM}

\begin{table*}
\tablenum{1}
\caption{1997 ASCA Observations of RCW 86 \label{tab:asca}}
\begin{center}
\begin{tabular}{lcccccccccc}
\tableline 
Name     &  RA (2000) &  DEC (2000) &
Sequence & Time     & Duration & Effective SIS \\
         & (hh mm ss) & ({}$^\circ$\ \ \arcmin\ \ \arcsec) &
Number   & (yy.ddd) & (ksec)   & Exposure (ksec)        \\
\tableline
RCW 86 (FRONT) &14 40 34.8 & -62 41 17 & 55045000 & 97.048 & 136 & 39\\
RCW 86 (BACK)  &14 40 54.9 & -62 38 48 & 55046000 & 97.050 & 121 & 42\\
RCW 86 (BK S1) &14 40 18.4 & -62 42 24 & 55046010 & 97.226 & 134 & 44\\
\tableline
\end{tabular}
\end{center}
\end{table*}

RCW 86 (G315.4--2.3) is a shell-like supernova remnant (SNR) in the
southern sky, 42\arcmin\ in angular size, and 35 pc in linear size at
a kinematic distance of 3 kpc found from optical observations by
Rosado et al. (1996). (The distance estimate is uncertain, and values
as low as 1 kpc cannot be excluded according to Smith 1997). Optical
emission from radiative shocks is present at various locations along
its shell, with by far the brightest complex of optical filaments in
the SW. This bright complex, to which the optical designation ``RCW
86'' actually refers, has been generally interpreted as the result of
the interaction of the blast wave with a dense ($\sim 10$ cm$^{-3}$)
interstellar cloud (e.~g., Rosado et al. 1996;
Smith 1997).  The slow ($\sim 100$ km s$^{-1}$) radiative shocks are
thought to have been driven into the cloud by the blast wave
impact. Optical, Balmer-dominated emission from faster nonradiative
shocks was also detected (Long \& Blair 1990), and Smith (1997) found
that Balmer-dominated optical filaments almost completely encircle RCW
86. This is presumably a blast wave propagating into a low-density
($\sim 0.2$ cm$^{-3}$, Long \& Blair 1990) interstellar medium
(ISM). Its velocity is in the range of 400--900 km s$^{-1}$, found
from the width of the broad H$\alpha$ component (Long \& Blair 1990;
Ghavamian 1999; Ghavamian et al. 2000). The remnant's age is $\sim 10^4$ yr,
derived from the remnant's angular size of 42\arcmin, its distance of
3 kpc and the known shock speed (Rosado et al. 1996). (However, note
that RCW 86 may be as young as 2000 yr if it is located at distance of
1 kpc.)  As expected, thermal X-ray emission produced behind fast
shocks (Pisarski, Helfand, \& Kahn 1984), nonthermal radio synchrotron 
emission from relativistic electrons
(Whiteoak \& Green 1996: Molonglo catalog), and strong IR
emission from collisionally heated dust (Dwek et al. 1987; Arendt 1989; 
Greidanus \& Strom 1990; Saken, Fesen, \& Shull 1992) are all present. 
RCW 86 was
thought to be a remnant of SN 185 (Clark \& Stephenson 1977), but even 
the existence of this SN has been questioned (Chin \& Huang 1994; 
Schaefer 1995).

RCW 86 is clearly an interesting SNR, seen across the wavelength band
from radio to X-rays, with both slow and fast shocks present. But the
most interesting is its peculiar X-ray spectrum as revealed by the
ASCA Performance Verification (PV) observations (Vink, Kaastra, \&
Bleeker 1997). The X-ray lines are generally much weaker than expected
from a normal (solar) abundance plasma, with the exception of the Fe
K$\alpha$ line at 6.4 keV. Vink et al. (1997) fitted ASCA data with
two-temperature nonequlibrium ionization (NEI) models and found
strongly subsolar abundances (except for Fe in their hot temperature
component) and extremely low ($\le 10^{10}$ cm$^{-3}$ s) ionization
timescales $n_et$. These results are clearly unphysical, and Vink at
al. speculated that perhaps they are caused by strong deviations from
the Maxwellian electron distribution for low-temperature plasmas.
A more recent study of RCW 86 by Bocchino et al. (2000), based on
BeppoSAX satellite data, confirmed the weakness of emission lines 
(other than the Fe K$\alpha$
line), although the authors attributed this effect to elemental depletion
onto dust. 

The goal of this paper is to explain the puzzling X-ray spectrum of
RCW 86, based on new, high signal-to-noise (S/N) ASCA observations. These new
observations are presented and briefly discussed in \S~2, while in \S~3
we analyze these observations in detail and show that they
can be succesfully explained as a combination of
thermal emission and nonthermal synchrotron emission.
In \S~4, we summarize our results and discuss prospects for further
progress.

\begin{figure}
\plotone{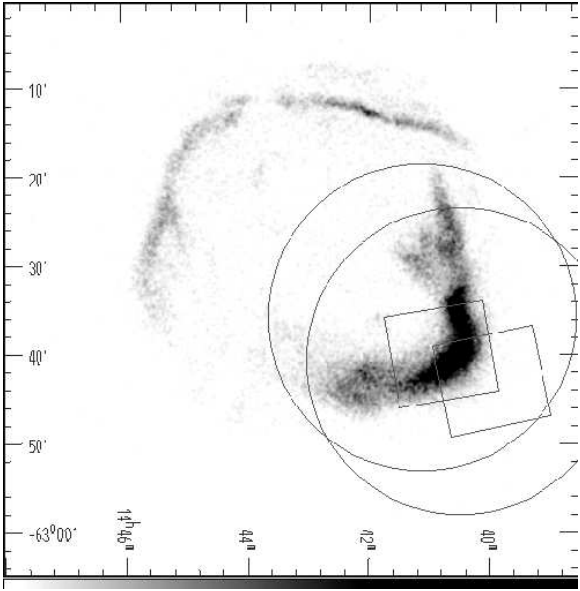}

\caption{ROSAT PSPC image of RCW 86. Fields 
of view of two ASCA observations from 1997 (SIS - squares, GIS - circles)
cover the brightest, SW ``knee'' region of the remnant.}
\label{rosimage}
\end{figure}

\begin{figure}
\plotone{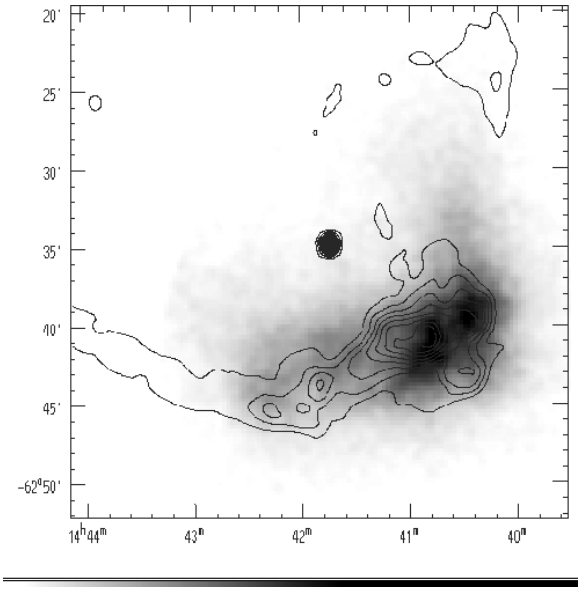}

\caption{GIS image of the SW knee of RCW 86, overlayed with 0.843 GHz radio 
contours (based on MOST radio data -- Whiteoak \& Green 1996). The X-ray and 
radio emission peaks nearly overlap, but radio
emission is much stronger in the horizontal arm to the west of the knee bend
than in the vertical arm to the north. }
\label{gisimage}
\end{figure}

\section{1997 ASCA OBSERVATIONS OF RCW 86 AND THEIR INTERPRETATION} 
\label{ascadata}

Three long observations of RCW 86 were performed by ASCA in 1997 
(Table~\ref{tab:asca}). The data were processed under standard
Revision 2 Data Processing (REV2) and screened according to the following 
criteria: elevation angle greater than 10 degrees, geomagnetic field structure
with cut-off rigidity greater than 6, observations outside South Atlantic 
Anomaly, stable pointing, and stable
internal background (from monitored Gas Imaging Spectrometers count rates). 
The effective exposure times after screening are typically about 40 ks as
shown in Table~\ref{tab:asca}. Note that the effective exposure time of the 
PV observations used by Vink et al. is only 14 ks. The long exposure times of 
the new observations resulted in good S/N spectra, leading to a significant
improvement over the PV observations, particularly at intermediate and high 
photon energies. This allowed us to separate thermal and nonthermal 
contributions to the X-ray spectrum.

\begin{figure}
\plotone{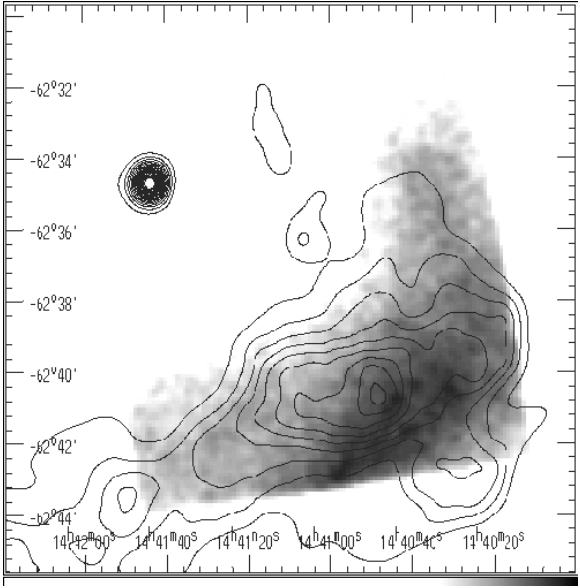}

\caption{SIS image of the SW knee of RCW 86, overlayed with 0.843 GHz radio  
contours. The bright, nearly horizontal ridge of 
radio emission is located interior of the brightest X-ray emission.}
\label{sisimage}
\end{figure}

\begin{figure}
\plotone{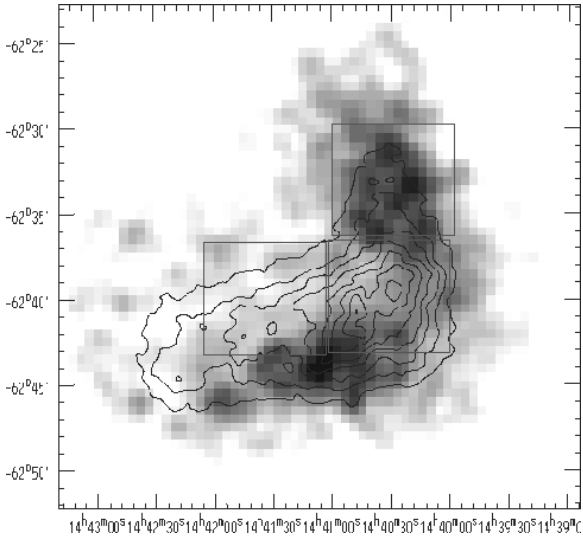}

\caption{The GIS hardness map, soft ($< 3$ keV) to hard ($> 3$ keV) count
rate ratio, with soft regions in black and hard regions in white, 
overlayed with GIS intensity  contours. Three SIS extraction regions are 
shown (clockwise, from the top: Soft
Region, Medium-Hard Region, and Hard Region).}
\label{sishmap}
\end{figure}

Fields of view of solid-state imaging spectrometers (SIS) and
gas-imaging spectrometers (GIS) for the first two observations listed
in Table~\ref{tab:asca} (sequences 55045000 and 55046010) are plotted on the 
ROSAT PSPC image of RCW 86
(Figure~\ref{rosimage}). The third pointing in the SW (sequence
55046000) is not shown, because its SIS field of view is centered outside
the remnant. However, this 
pointing contains useful GIS data on the SW part of the
remnant. These three long pointings provide high S/N images and
spectra of the brightest part of RCW 86. GIS and SIS images are shown
separately in Figures~\ref{gisimage} and~\ref{sisimage}, respectively,
while a GIS hardness map defined as the soft ($< 3$ keV) to hard ($> 3$ keV) 
count rate ratio is shown in Figure~\ref{sishmap}. The GIS hardness map 
is affected by the difference in the SIS and GIS point spread functions.
Significant spectral variations are readily apparent in this map, in
accord with findings of Vink et al.~(1997) which were based on PV
observations with a much lower S/N ratio. The bright ``knee'' in the
SW corner of the remnant exhibits striking variations in the hardness
ratio, with the soft vertical arm and the progressively harder
horizontal arm as one moves away (to the west) from the ``knee''
bend. The 0.843 GHz radio contours shown in Figures~\ref{gisimage}
and~\ref{sisimage}, from the Molonglo Observatory Synthesis Telescope
(MOST) catalog (Whiteoak \& Green 1996), demonstrate that the X-ray
hardness is positively correlated with the radio/X-ray brightness
ratio.

\begin{figure}[t]
\plotone{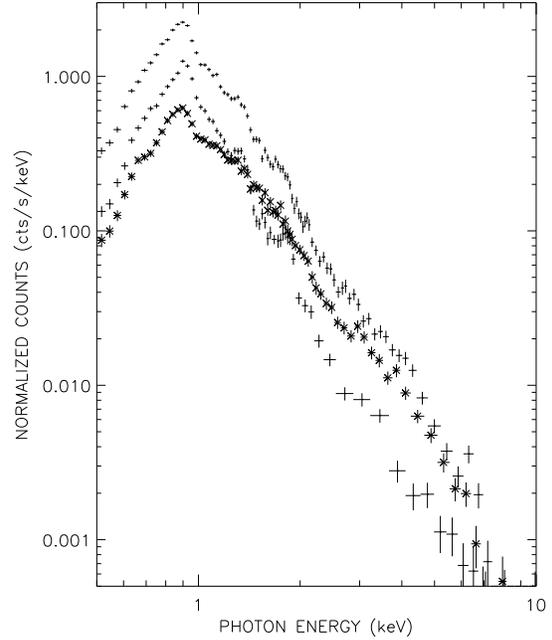}
\caption{SIS spectra for the three extraction regions (Medium-Hard -- top, 
Soft -- middle, Hard -- bottom, as seen below 1~keV). Weak emission lines are
clearly seen in all spectra. }
\label{sisspectra}
\end{figure}

SIS spectra (Figure~\ref{sisspectra}) of three rectangular regions
shown in Figure~\ref{sishmap} demonstrate spatial variations
quantitatively. Spectra of the vertical and horizontal arms, which we
label the Soft and the Hard Regions, respectively, differ
dramatically. The Soft Region is brighter at low energies, while the
situation reverses at high energies. The Medium-Hard Region at the
knee bend is intermediate in its hardness. Lines from O, Ne, and Mg
may be discerned in these spectra at low energies, although they are
apparently weak. Their presence indicates that X-ray spectrum of RCW
86 is at least partially of the thermal origin.

\begin{figure}[t]
\plotone{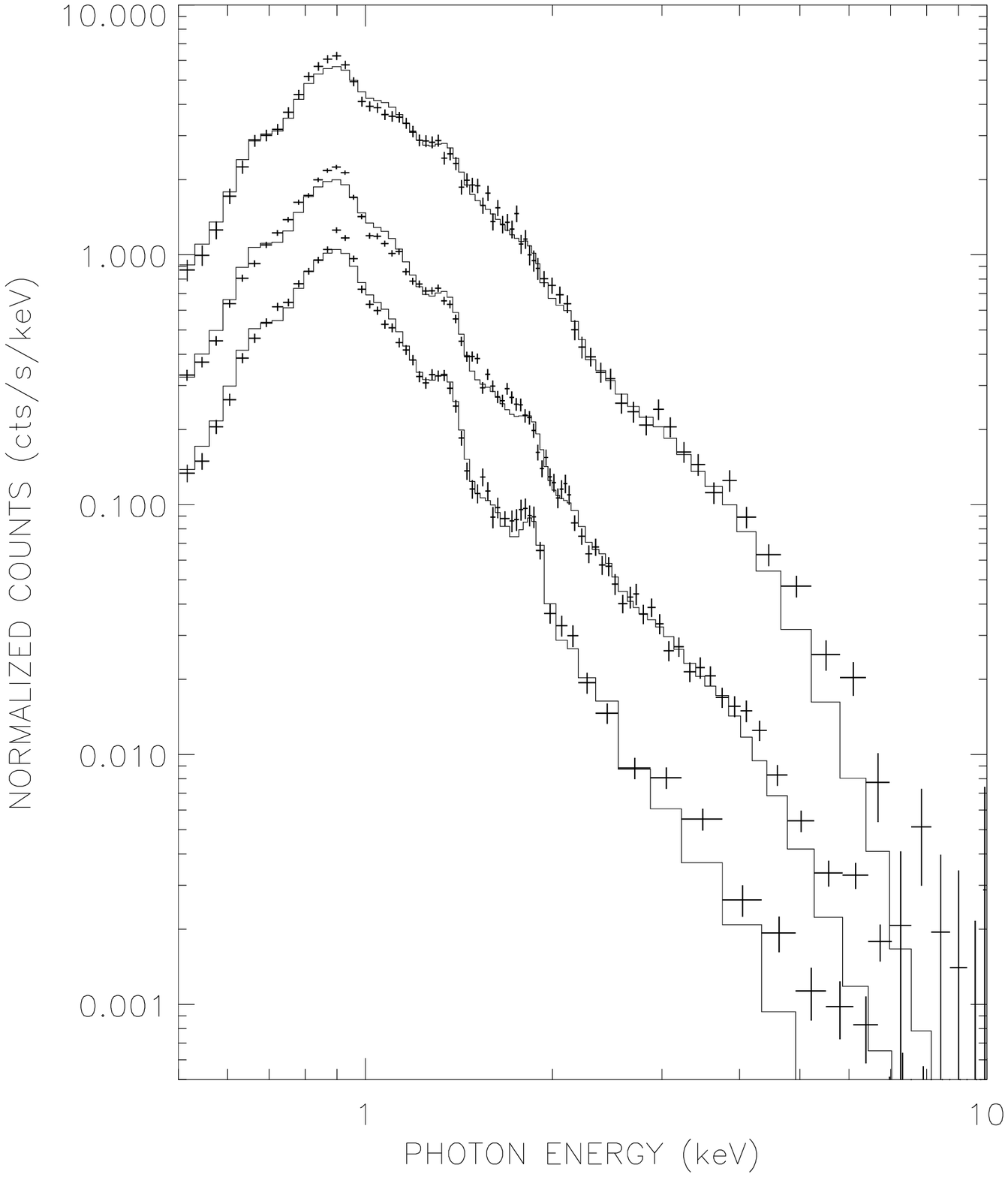}
\caption{Fits to SIS spectra with the {\it Sedov} models, shown as solid lines 
(Hard Region - top, 
Medium-Hard - middle, Soft - bottom). The spectrum of the Hard
Region was shifted upward by factor of 10 for the clarity of the
presentation. Metal abundances in {\it Sedov} models are low (0.2--0.55 solar)
and increasing from top to bottom.}
\label{sisthermal}
\end{figure}

We first test the hypothesis that X-ray spectra in
Figure~\ref{sisspectra} are produced entirely by thermal plasma, using
a set of nonequlibrium ionization (NEI) models developed by us and
publicly available in the XSPEC v11 software package. (For
introduction to XSPEC, see Arnaud 1996 or visit the web site at
http://heasarc.gsfc.nasa.gov.) Thermal NEI models are based on an updated 
Hamilton \& Sarazin (1984) spectral code, which includes recent atomic 
calculations for Fe L-shell lines by Liedahl, Osterheld, \& Goldstein (1995).
These models are discussed in more detail by Borkowski,
Lyerly, \& Reynolds (in preparation). We used {\it Sedov} models with the variable
metal abundances (but with their relative abundances fixed in solar
ratios)
to fit these spectra, with the best fit models shown in
Figure~\ref{sisthermal}.  Shock temperatures, postshock electron
temperatures, 
and SNR ionization timescales (equal to the product of
postshock electron density and the remnant's age) are: 1.4, 1.2, and
2 keV; 1.3, 1.2, and 0.7 keV;
and $5.0 \times 10^{10}$ cm$^{-3}$ s,
$5.4 \times 10^{10}$ cm$^{-3}$ s, and $8.9 \times 10^{10}$ cm$^{-3}$
s, for the Hard, Medium-Hard, and Soft Regions, respectively. In all
three regions we arrived at the absorbing column number density $N_H$
of $3 \times 10^{21}$ cm$^{-2}$.  Metal abundances in these models are
low (0.2, 0.4, and 0.55 solar for the Hard, Medium-Hard, and Soft
Regions, respectively), and spatially varying.  Spatial abundance
variations at the periphery of RCW 86 are not expected, and something
is obviously wrong with these models.  Fits with thermal models with
variable abundances presented by Vink et al. (1997) also lead to
strongly subsolar abundances of heavy elements and extremely low
ionization timescales. As these authors pointed out, this is clearly
unphysical, and they also concluded that standard thermal models do
not provide a satisfactory description of the X-ray spectrum of
RCW 86.

\begin{figure}[t]
\plotone{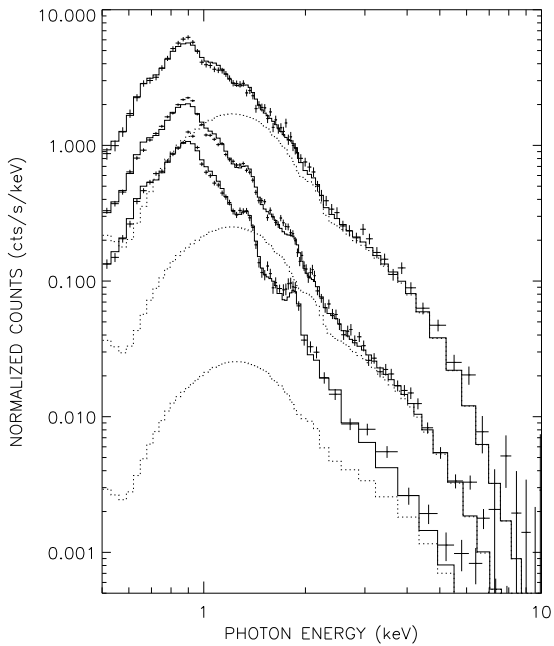}
\caption{Fits to SIS spectra with the sum of {\it Sedov} models and {\it srcut} nonthermal
synchrotron models. Nonthermal contribution, shown separately for each region, 
is most
important in the Hard region (top), and least important in the Soft region
(bottom). }
\label{sisthnth}
\end{figure}

The weakness of spectral lines suggests the presence of nonthermal
X-ray continuum in RCW 86, as has been recently inferred for SN 1006
\citep{Koyama95}, Cas A \citep{Allen97}, G347.5-0.2
\citep{Slane99}, and possibly even in RCW 86 itself based on its 
high-energy spectrum obtained by the {\it Rossi} X-ray Timing Explorer 
(Allen, Gotthelf, \& Petre 1999).  The spatial correlation between the radio 
and X-ray emission noted above suggests that this continuum may be
produced by synchrotron X-ray emission from relativistic electrons. In
order to test this hypothesis, we modeled SIS spectra as a sum of
nonthermal and thermal emission (plus interstellar absorption). We
used a simple nonthermal model with a power-law electron energy
distribution and an exponential cutoff at high electron energies (the
{\it srcut} model in XSPEC v11, see Reynolds 1998 and 
Reynolds \&\ Keohane 1999 for more details)
and a thermal {\it Sedov} model.  The {\it srcut} model is
parameterized by a 1 GHz flux density (in Jy) and spectral index, and a break
frequency which is the peak frequency emitted by electrons with the
$e$-folding energy of the exponential cutoff (at which the spectrum
has dropped by a factor of about 3.7).  Abundances were assumed to be
2/3 solar as appropriate for the ISM in the solar neighbourhood 
(e.~g., Snow \& Witt 1996; Mathis 1996).
Nonthermal models were normalized by the 1 GHz radio flux obtained by
integrating 0.843 GHz radio emission over the SIS extraction regions
and extrapolating to 1 GHz using the observed radio spectral index of
0.6 (Green 1998).  However, for the fitting, the
radio-to-X-ray spectral index was allowed to vary.  Resulting fits are
shown in Figure~\ref{sisthnth}. The quality of the fits is better than
for the pure thermal models, although they are still not statistically
acceptable. But unlike for pure thermal fits, parameter values of
these mixed thermal and nonthermal models are physically
reasonable. The {\it Sedov} model parameters are very similar in all three
regions: shock temperatures in the range from 1.0 to 1.4 keV, low ($<
0.2$ keV) postshock electron temperatures, and ionization timescales
in the range from $1.4 \times 10^{11}$ cm$^{-3}$ s to $1.8 \times
10^{11}$ cm$^{-3}$ s.  For the nonthermal {\it srcut} models we obtain
break frequencies of $7.5 \times 10^{16}$, $8.2 \times 10^{16}$, $2.5
\times 10^{17}$ Hz for the Hard, Medium-Hard, and Soft Regions,
respectively, and radio spectral
frequencies in the range from 0.62 to 0.67, in surprisingly good
agreement with the (uncertain) observed RCW 86 spectral index of 0.6.
The lower electron temperatures are a natural result of attributing
some of the continuum to nonthermal emission, but we stress that these
models are not well constrained and this may be an artifact.

\begin{figure}[t]
\plotone{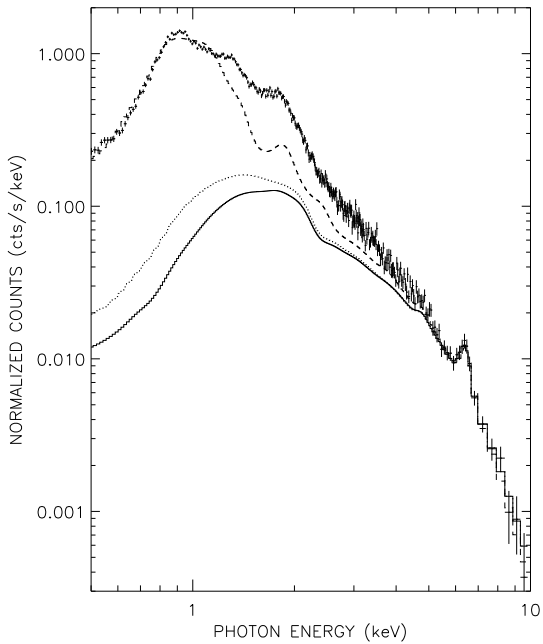}
\caption{The Fe K$\alpha$ line at 6.4 keV is clearly seen in the high S/N GIS2 
spectrum of the SW section of RCW 86. Three models providing acceptable fits to
energies above 5.5 keV are shown: a best-fit single ionization timescale model
{\it nei} with ionization timescale $\tau$ of $5 \times 10^8$ cm$^{-3}$ s 
({\it solid line}), an {\it nei} model with $\tau=6.7 \times 10^{9}$ cm$^{-3}$
s ({\it dashed line}), and a simple bremsstrahlung model with a Gaussian line
at 6.4 keV ({\it dotted line}).  }
\label{fekalpha}
\end{figure}

But the actual situation is more complex than described above, because
a weak Fe K$\alpha$ line at 6.4 keV seen in the SIS spectrum of the
Medium-Hard Region is not present in the models. This line, already
noted by Vink et al., is most likely produced by a high-temperature
thermal plasma. In order to investigate this line in more detail, we
combined available GIS2 data to construct a high S/N spectrum
(Figure~\ref{fekalpha}), which clearly shows Fe K$\alpha$ line. At
high energies above 5.5~keV, these data are fit well by a
bremsstrahlung continuum with electron temperature $kT_e = 5.3$ keV
(with 3.7--8.8 keV 90\% confidence interval) plus a narrow Gaussian
line at 6.42 keV (with 6.35--6.45 keV 90\% confidence interval). Such
a low energy of the Fe K$\alpha$ line suggests a very low ionization
timescale of the hot plasma. This is confirmed by fitting a single
ionization timescale, single temperature thermal NEI model ({\it vnei}
model in XSPEC, with variable Fe abundance and solar abundances for
other heavy elements) to our GIS data above 5.5~keV. We obtain plasma
temperature of 5.3 keV (with 3.6--8.6 keV 90\% confidence interval), a
very low ionization timescale of $5.2 \times 10^8$ cm$^{-3}$ s (less
than $6.7 \times 10^{9}$ cm$^{-3}$ s with 90\%\ confidence), and Fe
abundance of 3.5 solar (with 1.2--3000 90\% confidence interval).
Model parameters and confidence intervals depend quite sensitively on
the chosen energy range, resulting in progressively lower electron
temperatures as lower energy photons are included in the fits. Atomic
data for Fe K$\alpha$ line emission produced by electron excitation of
low-ionization stages of Fe are poor, further contributing to
uncertainties in derived parameter values, the Fe abundance in
particular. It is also possible that Fe K$\alpha$ is produced in dust
grains (Borkowski \& Szymkowiak 1997), which would make our estimates
of the ionization timescale unreliable. In this situation, the true
confidence intervals for all model parameters are certainly much
larger that the confidence intervals just quoted. In particular,
approximately solar abundance of Fe is consistent with the current
data.

The best-fit thermal bremsstrahlung and {\it nei} models produce
similar spectra across the GIS spectral range (Figure~\ref{fekalpha}),
without any prominent emission lines except for the fluorescent Fe
K$\alpha$ line. In the {\it nei} model this is caused by an extremely
low ionization state of the gas, where all heavy elements have not yet
been ionized to their H- and He-like stages (in the case of Fe even
M-shell electrons must still be present because of the apparent
absence of the Fe L-shell emission). The presence or absence of lines
depends very sensitively on the ionization timescale of the gas. If we
assume that the ionization timescale is equal to $6.7 \times 10^{9}$
cm$^{-3}$ s (the upper 90\%\ confidence limit in our fits), and fit
the {\it nei} model to the GIS2 data above 5.5~keV, the lines are
already very prominent (Figure~\ref{fekalpha}).  This extreme
sensitivity to the ionization timescale means that the contribution of
the high-energy thermal component to the X-ray spectrum at low photon
energies is not well constrained by the data, although in all cases
this contribution is not negligible and must be included in spectral
modeling.

The presence of a poorly-constrained high-temperature thermal
component in RCW 86 significantly complicates analysis of its X-ray
spectrum, and our previous arguments about the necessity of the
nonthermal component need to be reexamined. Is it possible that a
mixture of gas with standard ISM abundances but with widely different
temperatures results in the observed weakness of the lines? The answer
seems negative, according to Vink et al. (1997), and according to
Bocchino et al. (2000) who fitted BeppoSAX spectra of the SW region of
RCW 86 with a two-component NEI model. Bocchino et al. (2000) derived
the following parameters for the high-temperature component:
$kT_e = 5.7$ keV, ionization timescale of $1.9 \times 10^8$ cm$^{-3}$ s,
and the Fe abundance of 3.0 with respect to solar, in excellent agreement
with our results. But the derived abundances in the low-temperature
(1.0 keV) NEI component are again strongly (0.25) subsolar. We also
attempted to fit two-component ({\it Sedov} and {\it nei}) thermal models 
to our ASCA SIS spectra, but the
the derived abundances in {\it Sedov} models are
again strongly subsolar and spatially varying. Although this result does
not rule out the possibility of the pure thermal explanation, it
demonstrates that such an alternative is unlikely, in accord with Vink
et al. (1997). Another alternative explanation for the apparent subsolar
abundances, elemental depletion onto dust (Bocchino et al. 2000), is
not supported by the abundance patterns derived from ASCA and BeppoSax
observations. In view of these findings, we now present our efforts
to construct a physically reasonable model for the X-ray spectrum of
RCW 86, in which we allow for a simultaneous presence of nonthermal
synchrotron emission and thermal emission from multi-temperature
plasma.

\section{X-RAY SPECTRA OF RCW 86 AS A MIXTURE OF THERMAL AND NONTHERMAL 
EMISSION}

\begin{figure}[t]
\plotone{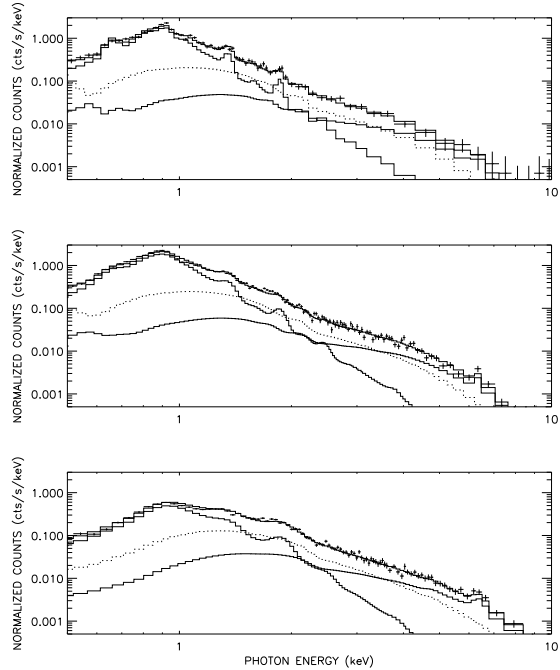}
\caption{ASCA spectrum of the Medium-Hard Region (PV SIS data -- top, 
1997 SIS data -- middle, GIS data -- bottom). A three-component 
({\it pshock} + {\it nei} + {\it scrut}) model is also plotted 
(total -- {\it thick solid line}, thermal components -- 
{\it thin solid lines}, nonthermal synchrotron component -- {\it dotted line}.}
\label{mediumhard}
\end{figure}

We begin with the analysis and modeling of the brightest extraction
region, the Medium-Hard Region at the knee bend (see
Figure~\ref{sishmap}).  In addition to the SIS spectra described in
the previous section, we also use the GIS spectrum extracted from the
same spatial region in order to improve S/N ratio at higher photon
energies. Below 1.2 keV, the PV SIS spectra are used instead of the
newer SIS and GIS data because of their superior spectral resolution
and good S/N ratio. These three data sets are shown in
Figure~\ref{mediumhard}. We always include these data sets together in
our analysis, by performing joint fits to the PV SIS data, and the
1997 SIS and GIS data.  In order to allow for mismatch between
normalizations of individual data sets, we allowed them to vary when
performing joint fits to the data.

The thermal plasma emission is modeled as a sum of low-temperature and
high-temperature components. For the low-temperature component, we use
either a simple plane-parallel shock model with constant electron
temperature $kT_e$ and ionization age $\tau$ (the {\it pshock} model
in XSPEC) or a much more complex spherically-symmetric shock with
unequal ion and electron temperatures (the {\it Sedov} model) which
was already described previously. The high-temperature component is
described
by a simplest NEI model, a single ionization timescale model {\it
nei}. The nonthermal synchrotron emission is modeled
by 
the {\it srcut} model, using the observed radio flux and its spectral index, 
and allowing the break frequency to vary. (If a power law is used
instead of the {\it scrut} model, the separation between the thermal and
nonthermal components becomes more difficult, and the fitted power-law index
and normalization of the power-law component are also difficult to interpret.)

Because of the presence of synchrotron emission, the chemical
abundances cannot be reliably determined from the observed spectra, so
that we assumed 2/3 solar abundances in the low-temperature thermal
component. Furthermore, the S/N ratio of the current data sets at high
photon energies is not adequate for a unique separation between the
high-temperature thermal component and the nonthermal synchrotron
emission. In this situation we assumed that all emission above 5.5 keV
is thermal, and used parameters of the high-temperature thermal
component obtained by fitting spatially-integrated GIS spectra above
5.5 keV (see \S~\ref{ascadata}). The normalization of this component
was obtained by fitting this thermal model alone to the spectral data
above 5.5 keV. The modeling results for the two models described above
are tabulated in Table~\ref{tab:models}.

We generally obtain similar results for our two types of models, {\it
pshock+nei+srcut} and {\it Sedov+nei+srcut}, although fits are always
better for the former. The best-fit {\it pshock+nei+srcut} model for
the Medium-Hard Region is shown in Figure~\ref{mediumhard}. The low-
(high-) temperature thermal component dominates at low (high)
energies, and the nonthermal synchrotron emission is most important in
the intermediate energy range. However, the nonthermal component
contributes at all energies, accounting for 22\%\ of the total
detected flux in the 0.5 -- 10 keV range (this fraction is shown in
the last row of Table~\ref{tab:models}). The top panel in
Figure~\ref{mediumhard}, with the high-quality PV SIS data, shows most
clearly how strong nonthermal emission leads to low equivalent widths
for the prominent Mg and Si lines.

\begin{table*}[t]
\tablenum{2}
\caption{Spectral Models \label{tab:models}}
\begin{center}
\begin{tabular}{lcccccc}
\tableline \tableline
          & \multicolumn{2}{c}{Medium-Hard Region} &
\multicolumn{2}{c}{Soft Region}  &
\multicolumn{2}{c}{Hard Region} \\
Model Parameters& A\tablenotemark{a} & B\tablenotemark{b} & A & B & A & B \\
\tableline
$N_H$ (cm$^{-2}$) & $1.9 \times 10^{21}$ & $2.5 \times 10^{21}$ & $2.8 \times 10^{21}$ & $3.3 \times 10^{21}$ & $3.2 \times
 10^{21}$ & $3.3 \times 10^{21}$ \\
$kT_l$ (keV)  & 0.75 & 0.90\tablenotemark{c} (0.04\tablenotemark{d} ) & 0.77 & 1.10 (0.07) & 0.80 & 1.4 (0.05\tablenotemark{e} )\\
$\tau_l$ (cm$^{-3}$s) & $1.8 \times 10^{11}$ & $3.1 \times 10^{11}$ & $1.1 \times 10^{11}$ & $2.2 \times 10^{11}$ 
& $5.7 \times 10^{10}$ & $1.5 \times 10^{11}$ \\
Abundance (fixed) & 2/3 & 2/3 &2/3 & 2/3 &2/3 & 2/3 \\
$EM_l/(4\pi d^2)$ (cm$^{-5}$)  & $9.8 \times 10^{11}$ & $1.3 \times 10^{12}$&
$1.1 \times 10^{12}$ & $1.4 \times 10^{12}$ & $4.2 \times 10^{11}$ & $5.3 \times 10^{11}$ \\
$kT_h$ (keV) (fixed) & \multicolumn{2}{c}{5.3} & \multicolumn{2}{c}{5.3} & \multicolumn{2}{c}{5.3} \\
$\tau_h$ (cm$^{-3}$s) (fixed)  & \multicolumn{2}{c}{$5.2 \times 10^8$} & \multicolumn{2}{c}{$5.2 \times 10^8$} 
& \multicolumn{2}{c}{$5.2 \times 10^8$} \\
Abundance (fixed) & \multicolumn{2}{c}{1} & \multicolumn{2}{c}{1} & \multicolumn{2}{c}{1} \\
Fe (fixed) & \multicolumn{2}{c}{3.4} & \multicolumn{2}{c}{3.4} & \multicolumn{2}{c}{3.4} \\
$EM_h/(4\pi d^2)$ (cm$^{-5}$)  & \multicolumn{2}{c}{$2.6 \times 10^{11}$} & \multicolumn{2}{c}{$1.2 \times 10^{11}$}
 & \multicolumn{2}{c}{$2.4 \times 10^{11}$} \\
$\nu_c$ (Hz) & $3.2 \times 10^{16}$ & $3.1 \times 10^{16}$ & $1.9 \times 10^{16}$ & $1.5 \times 10^{16}$ & $3.7 \times 10^{
16}$ & $3.7 \times 10^{16}$ \\
$\alpha$ (fixed) & 0.6 & 0.6 & 0.6 & 0.6 & 0.6 & 0.6 \\
$F_{1 GHz}$ (Jy) & 4.48 & 4.48 & 0.57 & 0.57 & 3.19 & 3.19 \\
X\tablenotemark{f} (0.5-10 keV) & 22\% & 20\% & 1.8\% & 1.2\% & 33\% & 33\% \\
\tableline
\end{tabular}
\end{center}
\tablenotetext{a}{$pshock+nei+srcut$}
\tablenotetext{b}{$Sedov+nei+srcut$}
\tablenotetext{c}{Postshock gas temperature}
\tablenotetext{d}{Postshock electron temperature}
\tablenotetext{e}{Fixed}
\tablenotetext{f}{Nonthermal/(nonthermal+thermal) flux ratio}
\end{table*}

\begin{figure}[t]
\plotone{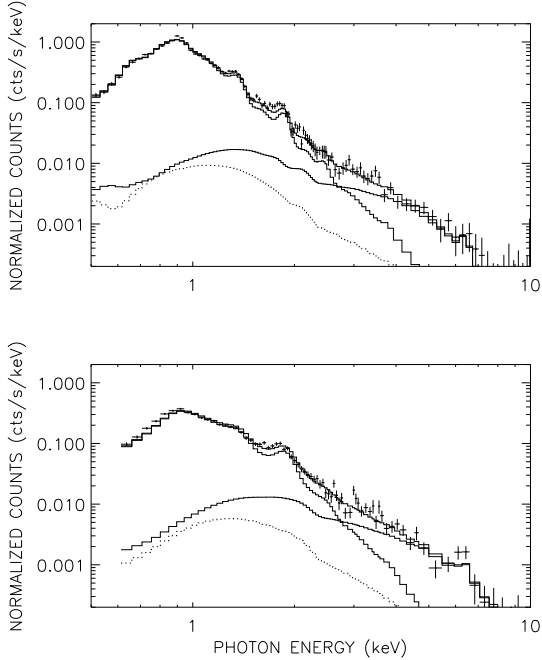}
\caption{ASCA spectrum of the Soft Region (SIS data -- top, 
GIS data -- bottom). A three-component model is also plotted (see Figure~9 for
explanation).}
\label{soft}
\end{figure}

Modeling of the Soft and Hard Regions is done in exactly the same way
as for the Medium-Hard Region, except that we do not use PV SIS data
because of their lower S/N. (For the Hard Region and the
{\it Sedov+nei+scrut} model, the shock temperature and the postshock
electron temperature in the {\it Sedov} model could not be determined 
independently, so that we fixed the postshock electron temperature
to a low value of 0.05~keV, similar to the postshock electron temperatures
found in the Medium-Hard and Soft Regions.)
Our SIS and GIS data for the Soft Region,
together with the best-fit {\it pshock+nei+srcut} model fom
Table~\ref{tab:models}, are shown in Figure~\ref{soft}. As expected,
the low- (high-) temperature thermal component dominates again at low
(high) energies, but the nonthermal component contributes only 1.8\% to
the total flux in this model. 

\begin{figure}[t]
\plotone{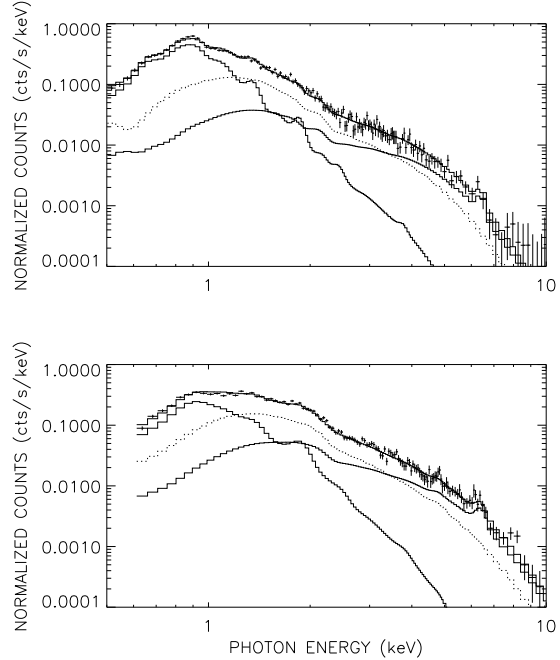}
\caption{ASCA spectrum of the Hard Region (SIS data -- top, 
GIS data -- bottom). A three-component model is also plotted (see Figure~9 for
explanation).}
\label{hard}
\end{figure}

The situation is very much different in the Hard Region
(Figure~\ref{hard}), where nonthermal emission is very prominent,
contributing 33\%\ of the total flux.
Again, emission at high energies is dominated by the
high-temperature thermal component with an Fe K$\alpha$ line clearly
seen at 6.4 keV, and faint lines with low equivalent widths at low
energies are produced by the low-temperature component. But the
emission measure of the latter is at least twice as low as in the
Medium-Hard and Soft Regions (see Table~\ref{tab:models}). At the same
time, the nonthermal component appears to be at the same intensity
level as in the Medium-Hard Region, leading to the increased fraction
of the nonthermal emission in the Hard Region.

The nonthermal synchrotron emission is strongly correlated with the
radio emission. In the {\it pshock+nei+srcut} model, after correction
for absorption, we obtain the total 0.5--10 keV X-ray synchrotron
fluxes of $9.5 \times 10^{-12}$ ergs cm$^{-2}$ s$^{-1}$, $6.1 \times
10^{-13}$ ergs cm$^{-2}$ s$^{-1}$, and $8.1 \times 10^{-12}$ ergs
cm$^{-2}$ s$^{-1}$ for the Medium-Hard, Soft, and the Hard Regions,
respectively. The corresponding 1~GHz radio fluxes are 4.5 Jy, 0.57
Jy, and 3.2 Jy. The low synchrotron X-ray flux is accompanied by the
low radio flux in the Soft Region, while an order of magnitude
increase in the nonthermal X-rays in the Medium-Hard and Hard Regions
is matched by a similar increase in the radio flux. Note that such
drastic changes are not seen in the thermal X-ray flux in the 0.5--10
keV range, which varies by only 20\% between the regions. Such strong
positive correlation between nonthermal X-ray and radio emission is
seen in the prototypical synchrotron-dominated remnant of SN1006. The
presence of this correlation in RCW 86 strongly supports the
synchrotron nature of the nonthermal emission in this remnant.

Our fitted values for the break frequency are all in the range
$\nu_c \sim (1.5 - 4) \times 10^{16}$ Hz. The energy of an
electron producing its peak synchrotron radiation at $\nu$ in a
magnetic field $B$ is given by
\begin{equation}
E_m = 45 \left( \nu_c \over {3.7 \times 10^{16} \ {\rm Hz}} \right)^{1/2}
      \left( B \over {10 \ \mu{\rm G}} \right)^{-1/2} \ {\rm erg}.
\end{equation}
Our values for $E_m$ then range from 14 to 45 erg (9 to 28 TeV), 
at the low end of the range of upper limits to $E_m$ found for bright
Galactic remnants by Reynolds \& Keohane (1999).  

Since RCW 86 is a more highly evolved remnant than other objects in
which synchrotron X-rays have been discovered, it is likely that the
mechanism causing the rolloff in the electron distribution above $E_m$
is synchrotron (and inverse-Compton) losses.  Reynolds (1998) gives
expressions for the maximum electron energy producible by shock
acceleration limited by radiative losses, assuming that the electron
mean free path is a factor $\eta$ times the gyroradius.  In the
strong-turbulence (Bohm) limit, $\eta \sim 1$.  For such conditions,
with a shock compression ratio of 4, we have
\begin{equation}
E_m \sim 50 \ \eta^{-1/2} \left( B_1 \over {3 \ \mu{\rm G}} \right)^{-1/2}
         u_8 \ {\rm erg}
\end{equation}
where $B_1$ is the upstream magnetic-field strength and $u_8$ the
shock velocity in units of $10^8$ cm s$^{-1}$.  Since
optical inferences (Ghavamian 1999; Ghavamian et al. 2000) give 
nonradiative-shock
velocities in the range of 400 to 900 km s$^{-1}$, we find predicted
values for the loss-limited break energy quite close to the
break energies we infer from the fits, for $B_1$ of a few microgauss
and $\eta \sim 1$.  A loss-limited spectrum
will not have the exact shape of the {\it srcut} spectrum, but
will roll off more slowly.  However, given the limited spectral
range of the fits, we cannot distinguish that difference, which in
any case is less than the difference between the {\it srcut}
spectrum and a straight power-law.  The coincidence between
expected and observed cutoff energies, while not an ironclad
conclusion, adds circumstantial support to the idea that
synchrotron X-rays are present in RCW 86.  The electron-acceleration
properties do not appear to vary much among the three regions
we have studied; the primary difference in the nonthermal
emission is just the overall strength of the emission, perhaps as
a result of differences in magnetic-field strength.  

The thermal X-ray emission consists of two distinct components: a
low-temperature component responsible for X-ray lines at low energies
and a high-temperature plasma where Fe K$\alpha$ line is produced. The
ratio of their emission measures systematically varies from a high
($\sim 10$) value in the Soft Region, through intermediate (4-5)
values in the Medium-Hard Region, and attains the lowest (2) value in
the Hard Region. The high- and low-temperature plasma should be in an
approximate pressure equilibrium, $n_{el}T_{el} \sim n_{eh}T_{eh}$,
where $n_{el}$ and $n_{eh}$ are electron densities in the low- and
high-temperature components, and $T_{el}$ and $T_{eh}$ are their
electron temperatures. For the Medium-Hard Region and the {\it
pshock+nei+srcut} model, $T_{el} = 0.75$ keV and $T_{eh} = 5.3$ keV,
and the assumption of a pressure balance gives a density ratio
$n_{el}/n_{eh} \sim 7$. But their emission measures differ by only a
factor of 4, which suggests that the filling fraction $f_h$ of the
high-temperature component is much higher that the filling fraction
$f_l$ of the low-temperature component. If we assume that the hot gas
fills most of the volume of the Medium-Hard Region, very approximately
equal to $6^3$ pc (where 6~pc is the linear size of the ASCA
extraction region at the assumed distance of 3~kpc), then we arrive at
$n_{eh} \sim 0.2~{\rm cm}^{-3}$ (where $f_h$ was set to 1), based on
the emission measure of the hot component listed in
Table~\ref{tab:models}.  Similarly, the low-temperature component has
density $n_{el} \sim 0.44/f_l^{1/2}$ cm$^{-3}$, while we obtain
$n_{el} = 1.6$ cm$^{-3}$ by assuming a pressure balance between the
high- and low-temperature components.  In order to make these
estimates consistent, a low filling fraction of 0.075 is required for
the low-temperature component. A low filling fraction $f_l$ is indeed
suggested by a highly filamentary morphology of this region of the
remnant seen in the archival HRI ROSAT images (Bocchino et al. 2000).
The thickness of
sharpest filaments seen in the HRI images is of the order of
10\arcsec\ or even less, which is much smaller than the 7\arcmin\
$\times$ 7\arcmin\ angular size of the Medium-Hard Region.

The low-temperature X-ray component is most likely produced in the
blast wave propagating into the ambient medium.  The electron
temperature $T_{el}$ of 0.8 keV in the {\it pshock+nei+srcut} models
implies a shock speed of 800 km s$^{-1}$ if electrons are fully
equilibrated with ions. In the absence of electron-ion equilibration,
the shock speed might be even higher, above 1000 km s$^{-1}$ as
suggested by mean postshock temperatures in excess of 1~keV in the
{\it Sedov+nei+srcut} models. The 800 km s$^{-1}$ shock velocity is at
the upper velocity range of nonradiative, Balmer dominated shocks seen
in RCW 86 (Long \& Blair 1990; Ghavamian 1999; Ghavamian et al. 2000).  
The preshock density $n_0 = n_{el}/4.8 \sim 0.3$ cm$^{-3}$ is also in accord 
with estimates of the preshock density in fast nonradiative shocks in
RCW 86. For example, Ghavamian (1999) finds preshock density of 1~cm$^{-3}$
for a $\sim 600$ km s$^{-1}$ shock located at the southern boundary of the
Hard and Medium-Hard Regions.
The shock age, $t \sim \tau_l/n_{el} = 3,600$ yr, is in reasonable
agreement with the remnant age estimated at $\sim 10^4$ yr by 
Rosado et al. (1996) from standard arguments based on the Sedov dynamics, and 
at 4,300 yr by Petruk (1999) whose estimates were based on 2-D hydrodynamical
simulations. This agreement further supports the identification of the
low-temperature thermal component with the material shocked by the
blast wave. Bocchino et al. (2000) also attributed the low-temperature
component to the shocked ambient gas based on analysis of BeppoSAX and
ROSAT data.

The high-temperature thermal component is produced by hot, tenuous gas
in the remnant's interior. The best evidence for this is provided by the
BeppoSAX hardness map (Bocchino et al. 2000), which shows that the hard
X-ray emission is located interior to the soft X-ray emission.
The high (5.3 keV) temperature of the hot component suggests that
this material was thermalized in fast, several thousand km s$^{-1}$
shocks at earlier stages of evolution of RCW 86. As determined by
fitting GIS data in \S~\ref{ascadata}, its ionization age is less than
$6.7 \times 10^9$ cm$^{-3}$ s at the 90\%\ confidence level, which in
combination with its electron density of $n_{eh} = 0.2$ cm$^{-3}$
leads to an upper limit of 1000 yr for its age. However, if Fe resided
in dust grains instead of gas, this upper limit would be several times
higher, and entirely consistent the blast wave age derived
previously. The Fe abundance in the hot gas is uncertain, but may be
enhanced with respect to the solar Fe abundance. If real, this would
imply the presence of heavy-element SN ejecta in the hot thermal
component, as already suggested by Bocchino et al. (2000). This component 
might have originated in the interaction of
the blast wave with a low-density cavity created by the SN progenitor
prior to its explosion (Vink et al. 1997).  It cannot be the normal
hot interior of a Sedov blast wave, at least for the Sedov
profile corresponding to the current blast-wave conditions, or
our previous Sedov fitting would have accounted for it.

Because of the simultaneous presence of two (or more) thermal components and 
of the nonthermal continuum, the separation of the X-ray spectrum of
RCW 86 into individual components is difficult and uncertain. The separation
of thermal emission into a low-temperature single shock component and a single 
ionization timescale component with high (and constant) temperature is
certainly a reasonable first approximation, but one might expect that such
a simple model would not be able to provide a perfect match to the observed
X-ray spectrum. This is indeed the case because our best fits are not
statistically acceptable at low photon energies, where ASCA provided a 
high-quality X-ray spectrum with the excellent S/N ratio. We speculate that
these problems stem from the simultaneous presence of shocks with different 
velocities in our extraction regions. Optical observations indicate presence
of shocks in a wide range of shock velocities from $\sim 100$ km s$^{-1}$ to
900 km s$^{-1}$ in various regions of RCW 86. Both slow 
($\sim 100$ km s$^{-1}$)
and fast ($\sim 800$ km s$^{-1}$) shocks are within the extraction regions,
as evidenced by the presence of radiative optical filaments and by the ASCA
spectra, respectively. While $\sim 100$ km s$^{-1}$ shocks cannot produce X-ray
emission, intermediate-velocity ($\sim 600$ km s$^{-1}$) shocks capable of
producing X-ray emission are present at the southern boundary of the 
Medium-Hard and Hard Regions 
(Long \&\ Blair 1990; Ghavamian 1999; Ghavamian et al. 2000). 
X-ray emission from such intermediate-velocity shocks might have been already
detected in the ROSAT PSPC spectra of RCW 86 by Bocchino et al. (2000), who
reported the presence of very low (0.29 keV) temperature gas in the SW region
of RCW 86. X-ray emission from these 
intermediate-velocity shocks is not accounted for in our simple decomposition
of the X-ray spectra. Our derived model parameters must then be considered
as preliminary rather than as robust estimates. In particular, low postshock
electron temperatures in {\it Sedov} models (Table~\ref{tab:models}) might be an
artifact of the model used instead of a measurement of the postshock electron
temperature. In view of the limited spatial and spectral resolution
of the ASCA satellite and its detectors, it would be premature at this time to 
construct a more sophisticated thermal models with multiple shock components.

\section{DISCUSSION}

Detailed modeling of X-ray spectrum of RCW 86 revealed the presence of
the nonthermal synchrotron emission, in addition to high- and
low-temperature thermal components. This nonthermal emission is
strongly correlated with the radio emission, and the derived break
frequency $\nu_c$ of $2-4 \times 10^{16}$ Hz is consistent with the
maximum electron energy producible by shock acceleration limited by
radiative losses. The presence of nonthermal synchrotron emission
explains why previous attempts to model X-ray spectrum of RCW 86 in
terms of pure thermal emission led to puzzling results.

The nonthermal synchrotron emission is most important at intermediate
(1--3 keV) photon energies, while at low (high) energies the ASCA
X-ray spectrum is dominated by a low- (high-) temperature thermal
component. This is a more complex behavior than seen in remnants such
as SN1006 where thermal emission dominates at low energies, while
nonthermal emission is most pronounced at high energies.  To model the
nonthermal component, a physically reasonable description is
necessary, as much for the implied inferences about the thermal
emission as for the synchrotron emission itself.  We do not expect
power-laws in X-ray synchrotron emission as we are always observing
the high-energy rolloff; using power-laws for modeling, while perhaps
allowed by the data in some cases, can lead to misleading inferences
about spectral-line equivalent widths.  For the {\it srcut} model with
a break frequency of $3.7 \times 10^{16}$ Hz, a power-law tangent to
that model at 2.1 keV (in the region where synchrotron emission
dominates in RCW 86) overpredicts the continuum at 6.4 keV by 50\%
compared to {\it srcut} -- so the equivalent width of the Fe K$\alpha$
line would be considerably underestimated by a power-law fit.

The nonthermal spectrum traverses the IR and optical on its way from
radio to X-ray wavelengths.  Since all the fitted break frequencies
are well above the frequency of the center of the $V$ band
($5.5 \times 10^{14}$ Hz), we can assume a straight extrapolation
from radio frequencies with a slope of $0.6$.  If the 4.5 Jy at 1 GHz
we attribute to the $7^\prime \times 7^\prime$ Medium-Hard region
were spread uniformly, they would imply a surface brightness
at $V$ of about 29$^{\rm m}$ arcsec$^{-2}$.  However, the morphology
should be identical to the radio; in particular, the radio emission
is concentrated in quite thin filaments, so that the peak brightness
is larger than average by at least a factor of 10, raising the
prediction to $26^{\rm m}.5$ arcsec$^{-2}$, perhaps attainable.  
At 2 microns, surface brightness would be almost a magnitude brighter.

The derived blast wave velocity of 800 km s$^{-1}$, the estimated
preshock density of $\sim 0.3$ cm$^{-3}$, and the shock age of $\sim 4
\times 10^3$ yr are consistent with previous estimates, although a low
(0.075) filling fraction for the gas shocked by the blast wave seems
to indicate that emission is dominated by denser than average
material. The enhanced X-ray emission and the presence of radiative
shocks in the bright SW section of the remnant have been attributed to
the interaction of the blast wave with a dense ($\sim 10$ cm$^{-3}$)
interstellar cloud.  Our findings support this interpretation. The
presence of a very hot ($\sim 5$ keV) gas with a large filling
fraction, with the emission measure far in excess of that expected in
a standard Sedov model, indicates that the SN progenitor most likely
exploded within a low-density cavity. This suggests that RCW 86 is a
remnant of a core-collapse SN, whose progenitor was perhaps a member
of an OB stellar association seen at a distance of 2.5 kpc in this
region of the sky (Westerlund 1969).

The presence of nonthermal synchrotron emission in RCW 86 is certainly
one of the most interesting aspects of this well-studied
remnant. Energetic electrons producing this emission have been
presumably accelerated in the blast wave, so it would be of great
interest to study in greater detail the relationship between the
thermal and nonthermal emission, and between optically-emitting shocks
and the nonthermal emission. The correlation of both thermal and
nonthermal X-ray emission with the strong far-IR emission from collisionally
heated dust grains seen in RCW 86
(Dwek et al. 1987; Arendt 1989;  Greidanus \& Strom 1990; Saken et al.
1992) is also important in view of the crucial role of dust grains in
acceleration of cosmic rays (Meyer, Drury, \& Ellison 1997; Ellison,
Drury, \& Meyer 1997). Such studies would allow for better
understanding of the shock acceleration process. Unfortunately, the
quality of the present X-ray observations does not allow for a unique
separation between the thermal and nonthermal emission. In order to
achieve this separation, future observations of RCW 86 with a spatial
and spectral resolution higher than that of ASCA are necessary, for
example with the XMM-{\it Newton} or {\it Chandra}
satellites. XMM-{\it Newton} should provide high S/N CCD spectra with
a moderate spatial resolution, while {\it Chandra} might be able to
resolve even the most compact X-ray filaments seen in the ROSAT HRI
images. The brightest X-ray filament seen in the ROSAT HRI image of
the SW region of RCW 86 (Bocchino et al. 2000) is of particular
interest, because it coincides with the brightest nonradiative shock
in RCW 86 whose shock velocity and the preshock density
was already determined from the optical spectroscopy (Ghavamian 1999).
It is also critical to obtain higher spectral resolution data,
at the resolution provided by the X-ray microcalorimeter onboard of
the failed Astro-E satellite. Such observations would allow for
determination of gas temperature from line diagnostics, and hence for
unambiguous separation of thermal and nonthermal components. Once
these components are separated, one can combine information from both
components to learn about acceleration of energetic (TeV) electrons in
SNRs. Maybe we can then find out why the nonthermal contribution is so
important in RCW 86, and relatively weak in some SNRs.

\acknowledgements We thank Parviz Ghavamian for providing us with useful
information about nonradiative shocks in RCW 86 in advance of
publication.  This research has made use of data obtained from the
High Energy Astrophysics Science Archive Research Center (HEASARC),
provided by NASA's Goddard Space Flight Center. We acknowledge the use
of MOST radio images of RCW 86. The MOST is operated by the University
of Sydney with support from the Australian Research Council and the
Science Foundation for Physics within the University of Sydney.
Support for this work was provided by NASA under grants NAG5-7406,
NAG5-7153, and NGTS-65.

\end{document}